\pdfoutput=1
\documentclass[10pt,conference,letterpaper]{IEEEtran}

\usepackage[noadjust]{cite}
\usepackage[T1]{fontenc}
\usepackage[utf8]{inputenc}
\usepackage{balance}
\usepackage{url}
\usepackage{subfig}
\usepackage{tabularx}
\usepackage{hhline}

\newcommand{\docauthor}{Jonathan Will, Jonathan Bader, and Lauritz Thamsen}
\newcommand{\docsubject}{Technische Universit\"at Berlin}
\newcommand{\dockeywords}{Scalable Data Analytics, Distributed Dataflows, Runtime Prediction, Resource Allocation, Cluster Management}
\newcommand{\doctitle}{Towards Collaborative Optimization of Cluster Configurations for Distributed Dataflow Jobs}

\PassOptionsToPackage{bookmarks=false}{hyperref}
\usepackage[pdftex]{hyperref}
\hypersetup{
    pdfborder={0 0 0},
    pdfauthor=\docauthor,
    pdftitle=\doctitle,
    pdfsubject=\docsubject,
    pdfkeywords=\dockeywords
}

\usepackage{amsmath,amssymb,amsfonts}
\usepackage{algorithmic}
\usepackage{graphicx}
\usepackage{textcomp}
\usepackage{xcolor}
\usepackage{tikz}

\newcommand\copyrighttext{%
  \footnotesize \textcopyright 2020 IEEE. Personal use of this material is permitted.
  Permission from IEEE must be obtained for all other uses, in any current or future
  media, including reprinting/republishing this material for advertising or promotional
  purposes, creating new collective works, for resale or redistribution to servers or
  lists, or reuse of any copyrighted component of this work in other works.
  DOI: \href{https://doi.org/10.1109/BigData50022.2020.9377994}{https://doi.org/10.1109/BigData50022.2020.9377994}}
\newcommand\copyrightnotice{%
\begin{tikzpicture}[remember picture,overlay]
\node[anchor=south,yshift=10pt] at (current page.south) {\fbox{\parbox{\dimexpr\textwidth-\fboxsep-\fboxrule\relax}{\copyrighttext}}};
\end{tikzpicture}%
}

\def\BibTeX{{\rm B\kern-.05em{\sc i\kern-.025em b}\kern-.08em T\kern-.1667em\lower.7ex\hbox{E}\kern-.125emX}}

\begin{document}

\title{\doctitle}

\author{%
\IEEEauthorblockN{\docauthor}
\IEEEauthorblockA{\docsubject, Germany\\
\{will, jonathan.bader, lauritz.thamsen\}@tu-berlin.de
}}

\maketitle
\copyrightnotice

\begin{abstract}
Analyzing large datasets with distributed dataflow systems requires the use of clusters.
Public cloud providers offer a large variety and quantity of resources that can be used for such clusters.
However, picking the appropriate resources in both type and number can often be challenging, as the selected configuration needs to match a distributed dataflow job's resource demands and access patterns.
A good cluster configuration avoids hardware bottlenecks and maximizes resource utilization, avoiding costly overprovisioning.

We propose a collaborative approach for finding optimal cluster configurations based on sharing and learning from historical runtime data of distributed dataflow jobs.
Collaboratively shared data can be utilized to predict runtimes of future job executions through the use of specialized regression models.
However, training prediction models on historical runtime data that were produced by different users and in diverse contexts requires the models to take these contexts into account.

\end{abstract}

\IEEEpeerreviewmaketitle

\begin{IEEEkeywords}
\dockeywords
\end{IEEEkeywords}

\section{Introduction}\label{sec:INTRO}
Distributed dataflow systems like Apache Spark~\cite{spark} and Flink~\cite{flink} make it easier for users to develop scalable data-parallel programs, reducing especially the need to implement parallelism and fault tolerance.
However, it is often not easy to select resources and configure clusters for executing such programs~\cite{perforator, Lama_AROMA_2012}.
This is the case especially for users who only infrequently run large-scale data processing jobs and without the help of systems operations staff.
For instance, today, many scientists have to analyze large amounts of data every now and again, in particular in areas like bioinformatics, geosciences, or physics ~\cite{bux2013parallelization},~\cite{pegasus_evo}.

In cloud environments, especially public clouds, there are several machine types with different hardware configurations available.
Therefore, users can select the most suitable machine type for their cluster nodes.
In addition, they can choose the horizontal scale-out, avoiding potential bottlenecks and significant over-provisioning for their workload.
Most users will also have expectations toward the runtime of their jobs.
However, predicting the performance of a distributed data-parallel job is difficult, and users often overprovision resources to meet their performance target, yet often at the cost of overheads that increase with larger scale-outs.

Many existing approaches in research iteratively search for suitable cluster configurations~\cite{cherrypick, hsu2018micky, hsu2018arrow, hsu2018scout}.
Several other approaches build runtime models, which are then used to evaluate possible configurations~\cite{ernest, aria, shah2019quick}, including our previous work~\cite{bell, ellis, verbitskiy2018cobell, koch2017smipe}.
Here, training data for the models is typically generated with dedicated profiling runs on reduced samples of the dataset.
Both approaches involve significant overhead for testing configurations.
This problem is aggravated in public cloud services like Amazon EMR that have cluster provisioning delays of seven or more minutes\footnote{\href{https://amzn.to/3rRbabd}{https://amzn.to/3rRbabd}, accessed October 20, 2020}.

Our previous work on cluster configuration additionally makes use of historical runtime data instead of relying on just dedicated profiling~\cite{bell, ellis, verbitskiy2018cobell, koch2017smipe}.
These approaches succeed in enterprise scenarios with many recurring workloads.

In a scientific context, monitoring data from previous executions are often not available, especially when resources for processing large datasets are only required relatively infrequently.
The sporadic nature of many data processing use cases makes using public clouds substantially cheaper when compared directly to investing in private cloud/cluster setups.

This presents an opportunity for collaboration since many different users and organizations use the same public cloud resources.
We expect especially researchers to be willing to share not just jobs, but also runtime metrics on the execution of jobs, in principle already providing a basis for performance modeling.\\

\vspace{-3mm}
\emph{Contributions}. The contributions of this paper are:
\begin{itemize}
    \item An idea for a system for collaboratively sharing runtime data to learn optimal cluster configurations for new distributed dataflow jobs
    \item A total of 930 unique runtime experiments\footnote{Available at \href{https://github.com/dos-group/c3o-experiments}{github.com/dos-group/c3o-experiments}} that are emulating executions from diverse collaborators across five commonly used distributed dataflow jobs
    \item A discussion of requirements for constructing runtime models that can work with heterogeneous historical runtime data\\
\end{itemize}
\vspace{-3mm}

\emph{Outline}. The remainder of the paper is structured as follows.
Section~\ref{sec:RELATED_WORK} discusses related work.
Section~\ref{sec:SYSTEM_IDEA} elaborates on the idea and proposes a system architecture for collaborative sharing of runtime data.
Section~\ref{sec:PRELIMINARY_RESULTS} presents the results of our experimental problem analysis.
Section~\ref{sec:REQUIREMENTS_FOR_RUNTIME_MODELS} discusses requirements for constructing suitable runtime prediction models.
Section~\ref{sec:CONCLUSION} concludes this paper and gives an outlook toward future work.

\section{Related Work}\label{sec:RELATED_WORK}
Our system aims to be applicable to more than one data processing system, which is why we devised a black-box approach for performance prediction.
This section consequently discusses related black-box approaches to runtime prediction and cluster configuration.

\subsection{Iterative Search-Based}

Some approaches configure the cluster iteratively through profiling runs, attempting to find a better configuration at each iteration, based on runtime information from prior iterations.
They finally settle on a near-optimal solution once it is expected that further searching will not lead to significant enough benefit to justify the incurred overhead~\cite{cherrypick, hsu2018micky, hsu2018arrow, hsu2018scout}.

For instance, CherryPick~\cite{cherrypick} tries to directly predict the optimal cluster configuration, which best meets the given runtime targets.
The search stops once it has found the optimal configuration with reasonable confidence. This process is based on Bayesian optimization.

Another example is Micky~\cite{hsu2018micky}.
It tries to reduce the profiling overhead by doing combined profiling for several workloads simultaneously.
For limiting overhead, it further reformulates the trade-off between spending time looking for a better configuration vs.\ using the currently best-known configuration as a multi-armed bandit problem.

Compared to these approaches, our solution avoids profiling and its associated overhead.

\subsection{Performance Model-Based}

Other approaches use runtime data to predict scale-out and runtime behavior of jobs.
This data is gained either from dedicated profiling or previous full executions~\cite{bell, ernest, verbitskiy2018cobell, ellis, koch2017smipe}.

For instance, Ernest~\cite{ernest} trains a parametric model for the scale-out behavior of jobs on the results of sample runs on reduced input data.
This works out well for programs exhibiting a rather straightforward scale-out behavior.
Ernest chooses configurations to try out based on optimal experiment design.

Another example is Bell~\cite{bell}, which includes a parametric model ba\-sed on that of Ernest and a non-pa\-ra\-met\-ric model with more accurate interpolation capabilities.
The system selects suitable data points to train its non-parametric model based on similarity to the current job, with the data points being taken from previous jobs.
Bell chooses between the two models automatically ba\-sed on cross-validation.
Additionally to profiling, it can learn the job's scale-out behavior from historical full executions, if those data are available.

The obvious disadvantage of all approaches based on dedicated profiling runs to gain training data is the associated overhead in both time and to some extent the cost.
Our proposed system will not rely on profiling runs.
Historical runtime data for a job is not always available within an organization.
We introduce a more comprehensive approach which can utilize runtime data that was generated globally and in vastly different contexts.

\section{System Idea}\label{sec:SYSTEM_IDEA}
This section presents our approach to the problem of finding the best cluster configuration for a distributed dataflow job.
We first present the overall concept and then explain a possible system architecture for an implementation of the approach.

\subsection{User Collaboration}

Especially with open source software, users share implementations of common jobs and algorithms instead of implementing these themselves.
Many of the most common distributed dataflow jobs are therefore being run every day by different individuals or organizations worldwide.
Consequently, the runtime data resulting from these executions could be shared for the benefit of all, enabling users to make accurate runtime predictions from the first execution of a job in their organization.
That is, the main idea of a collaborative optimization of cluster configurations is to share historical runtime data alongside the code for the jobs and prediction models, which allow users to benefit from global knowledge in both efficient algorithms and cluster configuration simultaneously.
Just like the users can contribute code to the repository in which they found the program they are using, they can also contribute their generated runtime data.

The code contributors to such repositories, henceforth called \textit{maintainers}, can use their domain knowledge to fine-tune the default models that come with the system to suit the job at hand or add entirely new, specialized models to it.

\begin{figure}[htbp]
    \centerline{\includegraphics[width=\columnwidth]{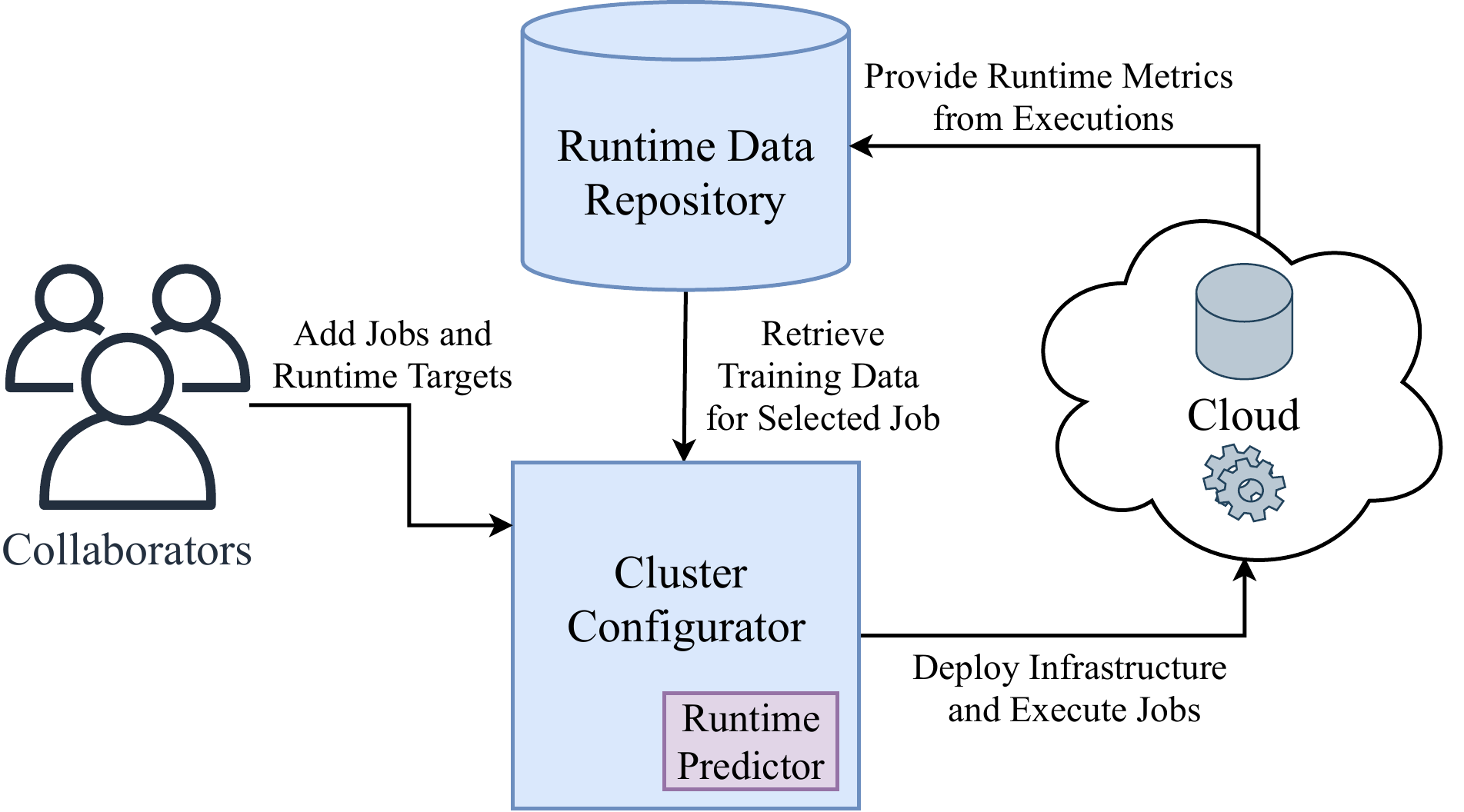}}
\caption{Workflow and user collaboration}\label{fig:collaboration}
\end{figure}

Fig.~\ref{fig:collaboration} illustrates the collaboration idea and depicts the envisioned workflow for the users.

\subsection{Envisioned Architecture}

Besides the code for the job, historical runtime data, and suitable runtime models, the system should contain a cloud interface to submit jobs and capture runtime statistics.
It is also used to extract key metadata about the user's dataset in cloud storage, like the size of the dataset in MB, which then serves as input to the runtime prediction models.
The exact configuration of this for a given job is the responsibility of the maintainers.

Finally, the repositories containing the code and the runtime data can be found by users on a website that lists them along with meta information, especially the underlying algorithm.

Altogether, the components form a system that streamlines the process of executing a distributed dataflow job on a user's data, as well as configure and create a cluster that fulfills the user's performance and budget constraints.

\begin{figure}[htbp]
    \centerline{\includegraphics[width=\columnwidth]{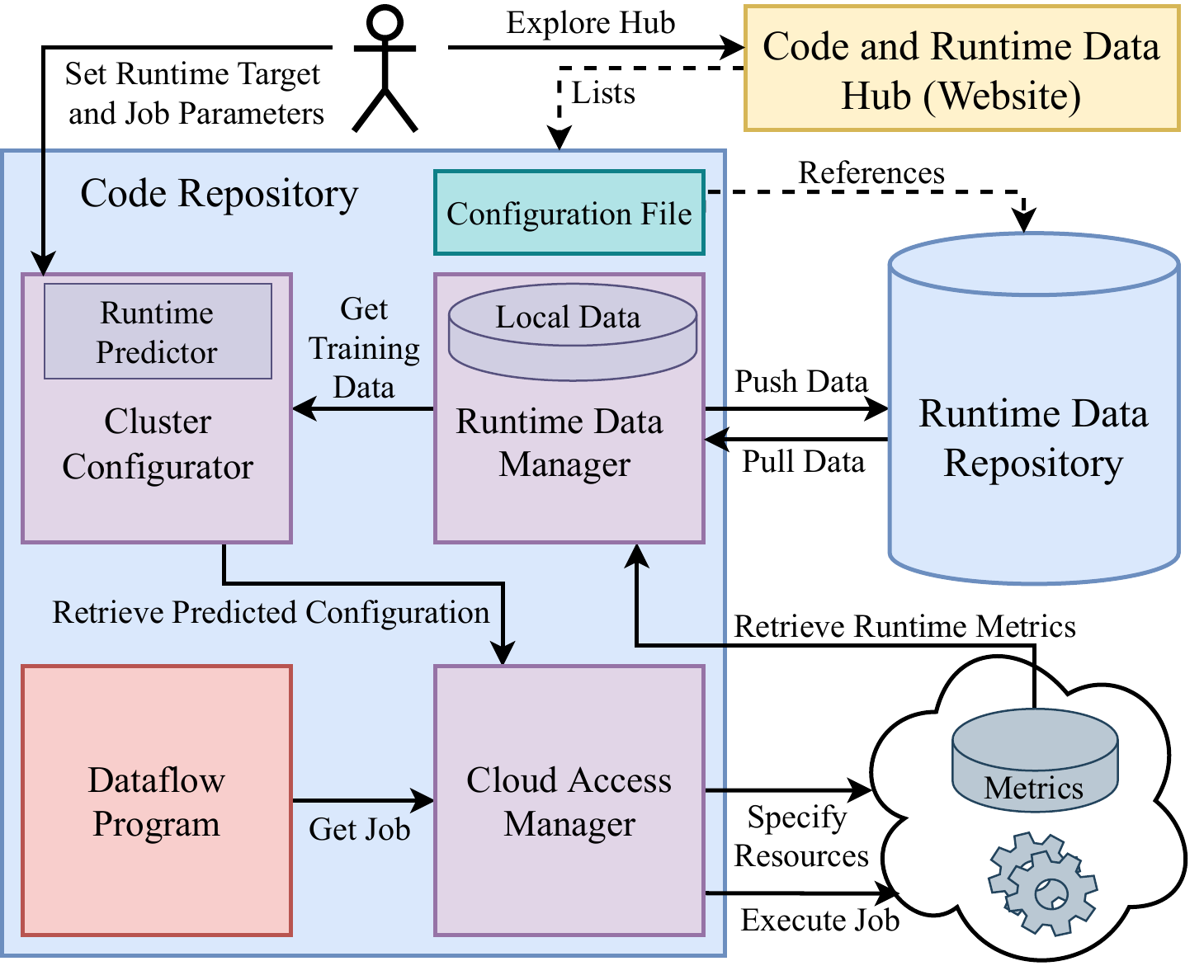}}
\caption{Overview of the envisioned system architecture}\label{fig:overview}
\end{figure}

Fig.~\ref{fig:overview} depicts the overall system architecture and its components.
This overview shows the \emph{code repository} and the \emph{runtime data repository} in light blue.
These are the two parts of the bundle that needs to be packaged for the user.
A system implementation contains exactly one fixed \textit{dataflow program}, shown in red, and the three modules shown in purple which contain default implementations.
Those can be adjusted by the maintainers to suit the given job more closely.

Users can provide job inputs in the form of a dataset location, parameters, and a runtime target, should one exist.
According to the runtime target, the \textit{cluster configurator} then uses training data retrieved by the \textit{runtime data manager} to predict the most suitable cluster configuration.
This is then reserved by the \textit{cloud access manager} and used to run the job.
Finally, the newly generated runtime data is captured and saved.

\subsection{Sharing Runtime Data}

The actual features to be collected and shared are determined by the maintainers.
Natural candidates are the job parameters and data characteristics, the latter of which might have to be determined during runtime, in case they are not clear from looking at a small subsample of the dataset.

In some cases, maintainers might decide to go with slightly alternative approaches, e.g.,\ for a Grep job.
Here, the input parameter \textit{keyword} alone does not deliver much useful information to runtime prediction models.
Instead, the number of occurrences of that keyword matters.

One way to implement sharing of historical runtime data alongside code is to put both into the same code repository.
A challenge here would be to prevent the code commit history from being diluted by data commits.

Another way to allow collaboration on runtime data from many users is to use a dedicated dataset version control system like DataHub~\cite{datahub} and reference it from the code repository.
An alternative is DVC\footnote{\href{https://dvc.org/}{dvc.org}, accessed October 20, 2020} which addresses code versioning and dataset versioning simultaneously.
Such systems provide functions like \textit{fork} and \textit{merge}, which are known from code version control systems.

If at some point, the dataset becomes too large for a quick download or fast training of prediction models, the overhead might grow too large to justify this approach over dedicated profiling.
A simple solution to this problem can be, however, to have the user only download a preselected sample of the historical runtime data of a specified maximal size, which covers the whole feature space most effectively.

\section{Preliminary Results}\label{sec:PRELIMINARY_RESULTS}
For a given data analytics workload, there is a multitude of factors that jointly determine the runtime of a job.
In the context of runtime prediction models, these factors can be referred to as features.
They include, for example, the specific distributed dataflow framework, the machine type and scale-out of the cluster, key dataset characteristics, and algorithm parameters.

Besides those, there are factors leading to runtime variance that are rather difficult to predict and account for, e.g.,\ partial or complete system failures and subsequent recovery routines.
Therefore, we do not attempt to model them.

In this section, we examine how to construct a suitable runtime predictor.
This predictor lies at the core of our collaborative cluster configuration solution.
It must be well-adapted to the particularities of runtime data from users in different contexts, such as having vastly different scale-outs or dataset sizes.

\subsection{Experimental Setup}

Five different algorithms were tested under various cluster configurations in Amazon EMR 6.0.0, which uses Hadoop 3.2.1 and Spark 2.4.4.
The JAR files containing those algorithms were compiled with Scala version 2.12.8

In total, we executed 930 unique runtime experiments, an overview of which can be seen in Table~\ref{table:test_jobs}.
Each of the runtime experiments was conducted five times, and the median runtimes are reported here in order to control for outliers.
We used the standard implementations that come with the official libraries of Spark for the algorithms.

\begin{table}[h]
    \centering
    \caption{Overview of Benchmark Jobs}
    \begin{tabularx}{\columnwidth}{|l|c|X|c|X|}
    \hline
             & Jobs & Datasets                            & Input Sizes & Parameters                                         \\ \hhline{|=|=|=|=|=|}
    Sort     & 126  & Lines of random chars               & 10-20~GB    & ---                                                \\ \hline
    Grep     & 162  & Lines of random chars and keywords  & 10-20~GB    & Keyword\newline ``Computer''                       \\ \hline
    SGD      & 180  & Labeled Points                      & 10-30~GB    & Max.\ iterations 1-100                             \\ \hline
    K-Means  & 180  & Points                              & 10-20~GB    & 3-9 clusters,\newline convergence criterion 0.001  \\ \hline
    PageRank & 282  & Graph                               & 130-440~MB  & convergence criterion\newline 0.01-0.0001          \\ \hline
    \end{tabularx}
    \label{table:test_jobs}
\end{table}

\subsection{Experiments}

In the following, we show a selection of the most important experiment results.

\subsubsection{Machine Type Selection}\label{subsubsec:machine-type-selection}

One objective of cluster configuration is to find the most resource-efficient machine type for the problem at hand.
Different algorithms have different resource needs regarding CPU, memory, disk I/O, network communication.
Naturally, an efficient machine fulfills all those needs, avoiding hardware bottlenecks.

\begin{figure}[htbp]
    \centerline{\includegraphics[width=\columnwidth]{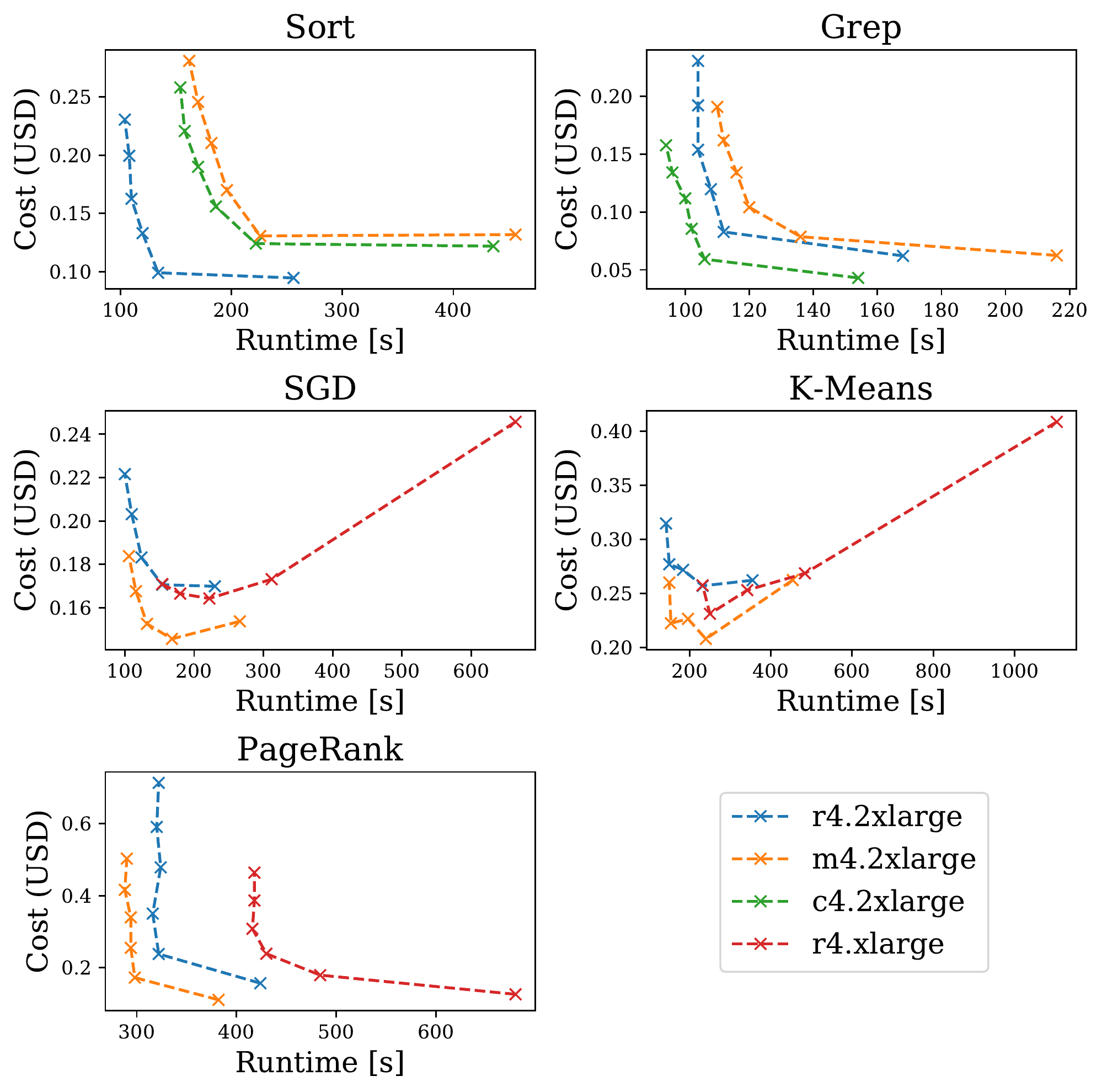}}
    \caption{Machine types and cost-efficiency at different\\scale-outs. Instance count left to right: 12, 10, \ldots}\label{fig:mtypes}
\end{figure}

Fig.~\ref{fig:mtypes} shows how the cost-efficiency of various machine types behaves at different scale-outs.
Lower scale-outs naturally come with long runtimes.
Typically they also lead to lower costs.
Exceptions to this rule are memory bottlenecks that can occur at lower scale-outs, which can also be seen in the cases of SGD and K-Means.
This phenomenon has also been noted in related work~\cite{hsu2018arrow}.

Our main conclusion from the results in Fig.~\ref{fig:mtypes} is that the cost-efficiency ranking of machine types should remain mostly static for a given algorithm, even throughout different scale-outs.
Thus, the machine type choice can generally be made based solely on the data analytics algorithm at hand and should be largely independent of the algorithm's inputs or the user's runtime requirements.
This observation is in accordance with related work~\cite{hsu2018micky}.

\subsubsection{Dataset Characteristics}

\begin{figure}[htbp]
    \centerline{\includegraphics[width=\columnwidth]{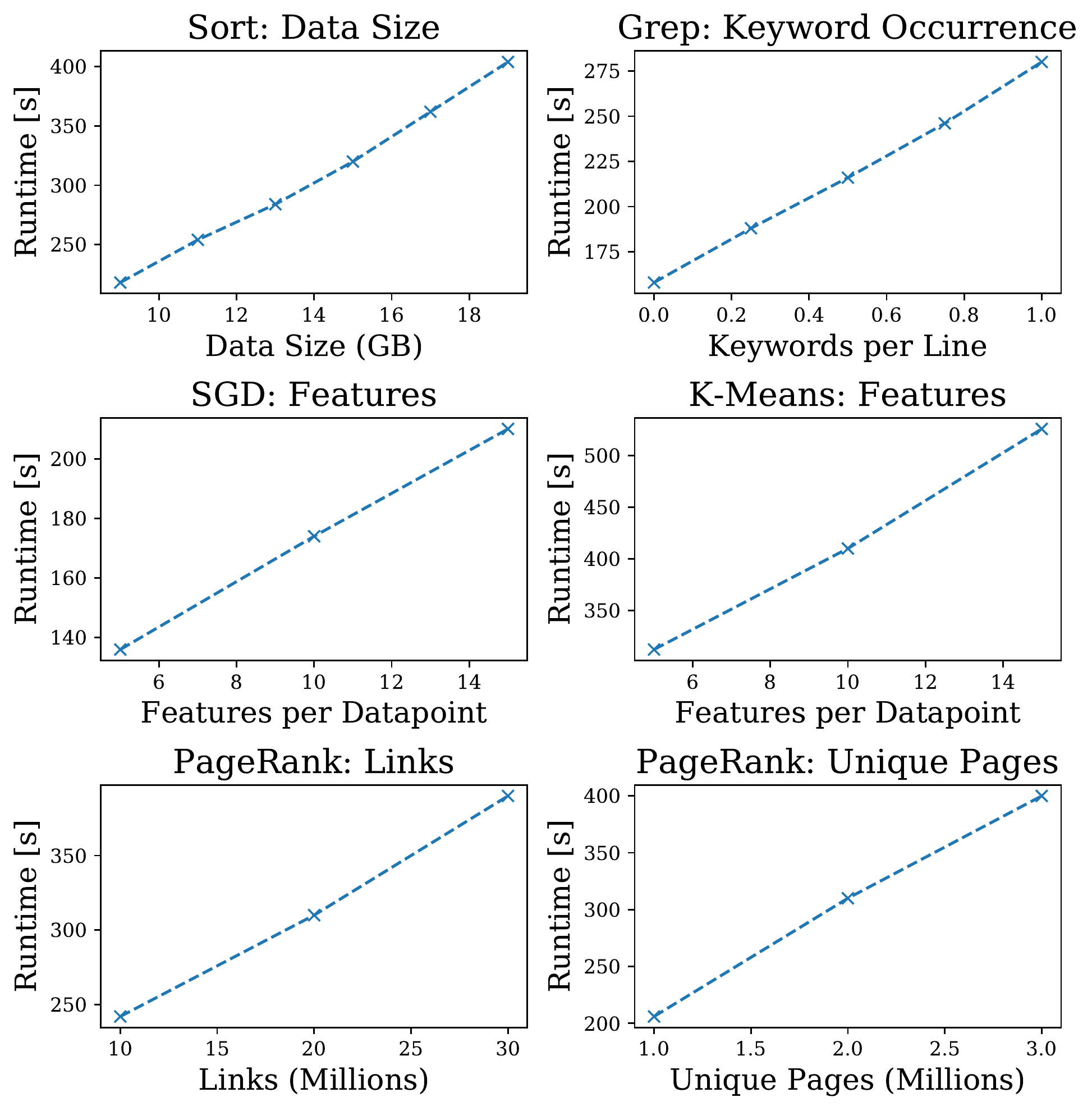}}
\caption{Influence of key data characteristics on the runtime}\label{fig:problem_size}
\end{figure}

Fig.~\ref{fig:problem_size} shows how key data characteristics influence the runtime of the data analytics workloads tested experimentally.
Aside from the ones examined here, all other runtime-influencing factors for each of the algorithms remained fixed.
The examined data characteristics appear to influence the problem size, and therefore the runtime linearly.

\subsubsection{Algorithm Parameters}

\begin{figure}[hbtp]
    \centerline{\includegraphics[width=\columnwidth]{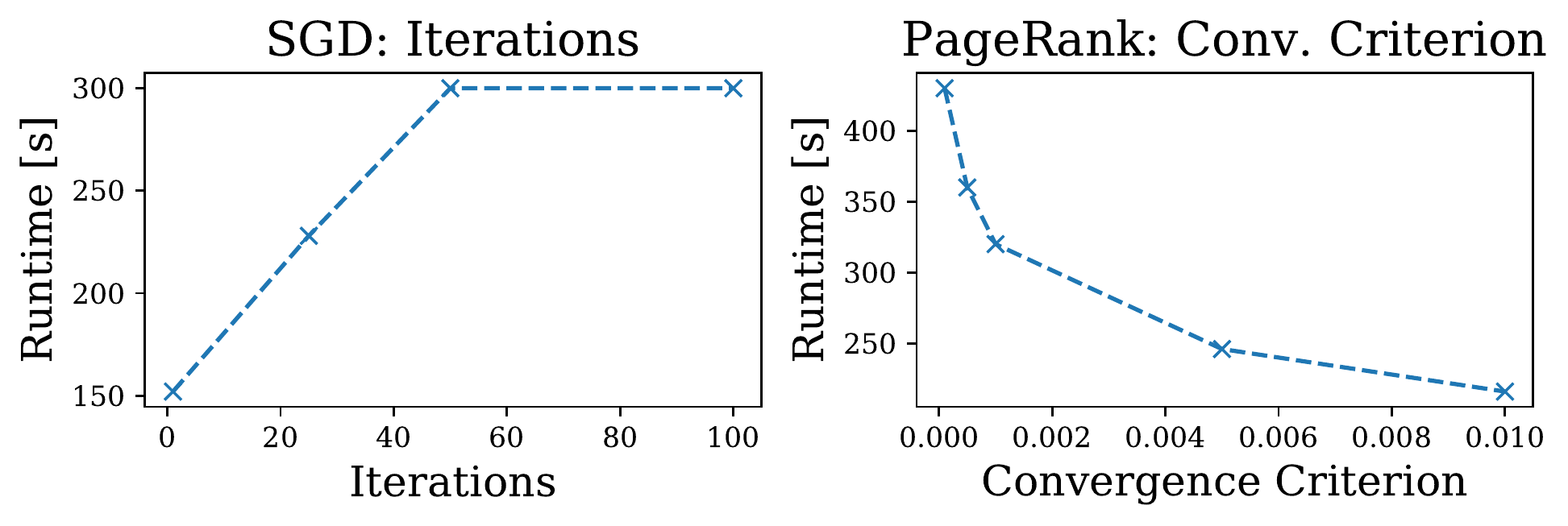}}
\caption{Influence of different input parameters on the runtime}\label{fig:parameters}
\end{figure}

Fig.~\ref{fig:parameters} shows the influence of a selection of algorithm parameters on the runtime of tested data analytics workloads.
Again, aside from the ones examined here, all other runtime-influencing factors for each of the algorithms remained fixed.
The ones examined influence the runtime of the respective workload non-linearly.

\subsubsection{Scale-Out Behavior}

Different parallel programs can have vastly different scale-out behavior.
Besides that, the implementation of the program and the distributed dataflow framework also have an influence.

\begin{figure}[htbp]
    \centerline{\includegraphics[width=\columnwidth]{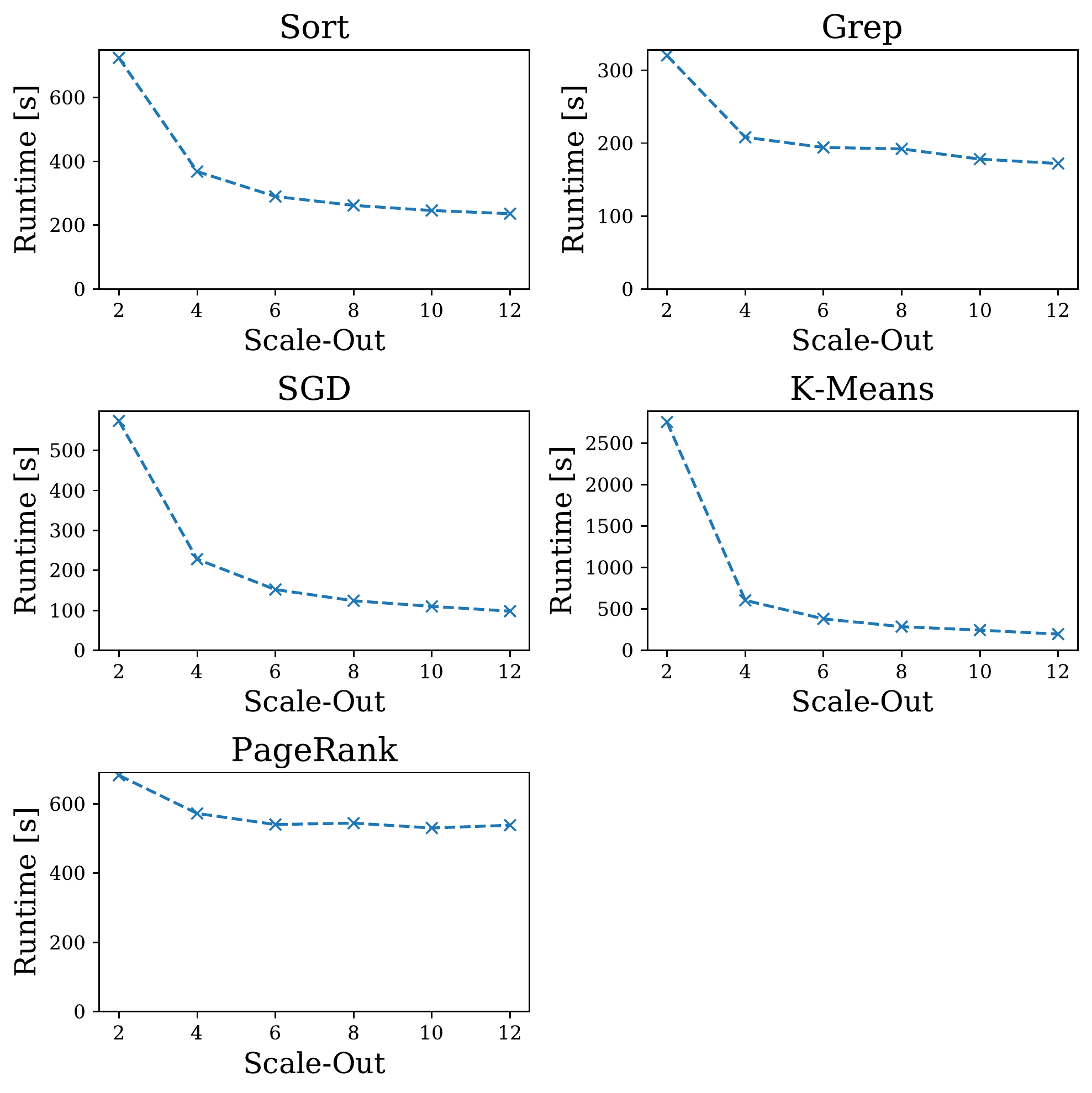}}
\caption{Scale-out behavior}\label{fig:scaleout}
\end{figure}

Fig.~\ref{fig:scaleout} shows the scale-out behavior for the tested algorithms.

Again, it is visible what we believe to be memory bottlenecks in SGD and K-Means, which in both cases appear to occur at a scale-out of two.
Doubling the node count from two to four leads to speed-up $> 2$ in both cases.

Another noteworthy detail is that PageRank appears to benefit relatively little from scaling out.

\begin{figure}[htbp]
    \centerline{\includegraphics[width=\columnwidth]{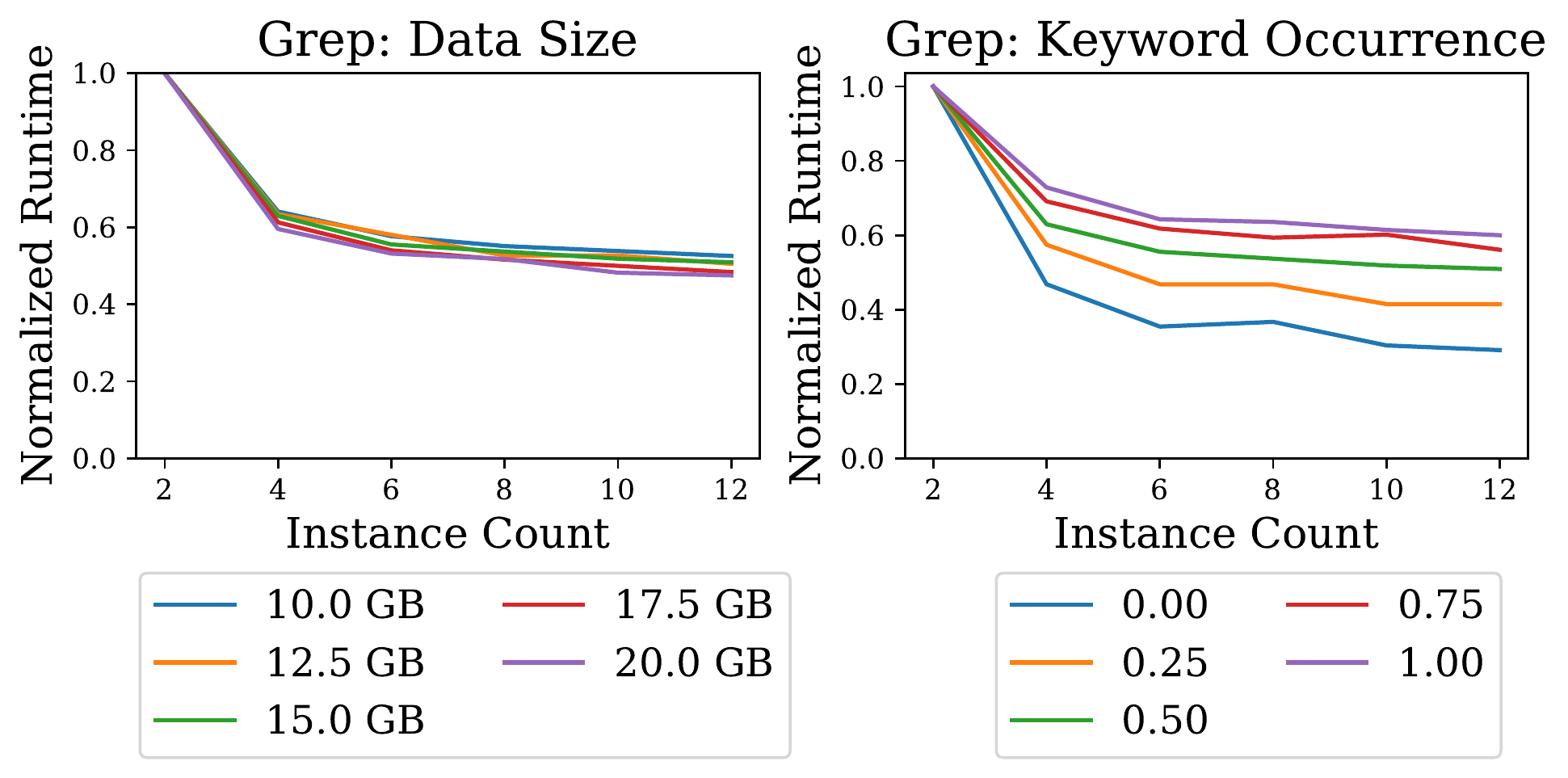}}
\caption{Scale-out behavior vs other factors}\label{fig:grep_scaleout}
\end{figure}

Exemplified in Fig.~\ref{fig:grep_scaleout} is the influence that input data characteristics can have on the scale-out behavior of a data analytics workload.
In the case of Grep, one can see that the size of the dataset does not significantly influence the scale-out behavior, while the ratio of lines containing the keyword does have an influence.

Looking through the file for keywords can be done in parallel.
The algorithm then writes lines with the found keyword back to disk in their original order, which is done sequentially.
It demonstrates that certain data characteristics, in this case the occurrence ratio of the keyword, can change the scale-out behavior of the job.

While Fig.~\ref{fig:grep_scaleout} shows only two examples, our remaining analysis on this matter can be summarized as follows.

In our examination, nearly every runtime-contributing factor mentioned in this chapter does \textit{not} significantly influence the scale-out behavior of its respective job.

\section{Requirements for Runtime Models Learning from Shared Training Data}\label{sec:REQUIREMENTS_FOR_RUNTIME_MODELS}
In a collaborative setting, runtime metrics produced globally by different users can be expected to vary in all the previously highlighted runtime influencing factors.
The exact amount of features that go into the prediction models depends on the algorithm and how many input parameters and key data characteristics are considered.
In any case, however, they are numerous.

An increasingly high feature space dimensionality renders available training data sparse.
Naturally, the training dataset for the runtime prediction algorithms being sparse complicates making accurate predictions.
One way to counter this and to see an increased prediction accuracy is by having more training data, but the effectiveness of that is down to circumstance.

For designing models that can cope with this high feature space dimensionality in our case, we have identified two generalizable approaches. They are introduced in the following.

\subsection{Pessimistic Approach}

We can make use of the fact that many jobs are recurring, at least within an organization~\cite{jyothi2016morpheus,agarwahl2012reoptimizing}.

It would be fair to assume that between recurring executions of a job most algorithm parameters and some key data characteristics should not change, only perhaps the problem size.
Predictions with this approach are made based on the most similar previous executions.
Similarity can be assessed by finding appropriate distance measures in feature space and scaling each feature's relative distance by that feature's correlation with the runtime.
Estimating runtimes from configurations that are equal or near-equal to historical configurations is therefore enabled especially by this approach.
It succeeds almost regardless of feature-dimensionality and interdependence.

\subsection{Optimistic Approach}

This approach optimistically assumes that the features influence the runtime of the job independently of one another.
In our experimental problem analysis, that assumption holds in most cases, meaning that most features are pairwise independent.

Thus, the strategy is to learn the influence of (groups of) pairwise independent features and then finally recombine those models.
This results in several models of low-dimensional feature spaces. Owing to the curse of dimensionality as described by Richard E. Bellman, these together require less dense training data than single models that consider all features simultaneously.

\subsection{Dynamic Model Selection}

Which of these approaches performs better depends on the particular situation.
The dataflow job and the specific implementation of the respective models  influence the accuracy of predictions.
Also, the quantity and quality of available training data points are important factors to be considered.
Models based on the pessimistic approach are expected to perform well on interpolation, when dense training data is available, or on recurring jobs.
Conversely, the optimistic approach-based models are expected to have better extrapolation capabilities even with relatively sparse training data, given mostly independent features.

Training data characteristics change as time progresses and more training data become available.
Hence, we intend to switch dynamically between prediction models depending on expected accuracy.
The models are retrained on the arrival of new runtime data.
Based on cross-validation, the most accurate model averaged over the test datasets is chosen to predict new data points.

\section{Conclusion and Future Work}\label{sec:CONCLUSION}
The goal of this work is to design a new system that is capable of configuring an efficient public cloud cluster for data analytics workloads while fulfilling the users' runtime requirements.
Towards this goal, we designed a collaborative system that allows users to share historical runtime data of distributed dataflow jobs.
The runtime data is shared alongside the code of the job and is used to train black-box runtime prediction models which lie at the core of our cluster configuration system.
Our prediction models need to cope with the high dimensionality of predicting performance based on historical executions of jobs in different organizations by making use of the characteristics of jobs and runtime data.
The runtime predictor of our envisioned system switches dynamically between a selection of suitable runtime prediction models based on expected accuracy in a given situation.

In the future, we want to work on effective runtime prediction models based on both approaches outlined as well as strategies for adaptively switching between multiple prediction models.
Moreover, we are working on a prototype for the entire collaboration system, which we want to make publicly available.

\section*{Acknowledgments}

This work has been supported through grants by the German Ministry for Education and Research (BMBF) as BIFOLD (grant 01IS18025A) and the German Research Foundation (DFG) as FONDA (DFG Collaborative Research Center 1404).

\bibliographystyle{IEEEtran}
\balance
\bibliography{./references}

\end{document}